\documentclass[english]{achemso}
\usepackage[T1]{fontenc}
\usepackage[latin9]{inputenc}
\usepackage{units}
\usepackage{amsmath}
\usepackage{graphicx}
\usepackage{esint}

\makeatletter

\title{Uncertainty and auto-correlation in Measurement}

\author{Markus Schiebl}

\affiliation{Federal Office of Metrology and Surveying (BEV), 1160 Vienna, Austria}

\email{markus.schiebl@bev.gv.at}

\numberwithin{equation}{section}
\numberwithin{figure}{section}

\makeatother

\usepackage{babel}
\begin{document}
\begin{abstract}
Although a system is described by a well-known set of equations leading
to a deterministic behavior, in the real world the value of a measurand
obtained by an experiment will mostly scatter. Accordingly, an uncertainty
is associated with that value of the measurand due to apparently random
fluctuation. This papers deals with the question why this discrepancy
exist. Furthermore it will be shown how the uncertainty of one individual
observation is calculated and consequently how the best estimate and
its corresponding uncertainty considering auto-correlations is determined. 
\end{abstract}

\section{Introduction}

A measurand is determined from other quantities trough a functional
relationship $f$ by \cite{key-1},

\begin{equation}
y=f\left(x_{1},x_{2},...,x_{n}\right)\label{eq:1.1}
\end{equation}

where $x_{1},x_{2},...,x_{n}$ are input parameters. These quantities
are often in turn influenced by other quantities trough a functional
relationship $g$ by,

\begin{equation}
x_{i}=g_{i}\left(\varepsilon_{1},\varepsilon_{2},..,\varepsilon_{k}\right)\label{eq:1.2}
\end{equation}

where the parameters $\varepsilon_{1},\varepsilon_{2},..,\varepsilon_{k}$
are fundamental since they determine the characteristics of the measurand
via Eq. (\ref{eq:1.1},\ref{eq:1.2}).

Unfortunately, not all fundamental parameters may be known , e.g.,
convection of air causing dynamic pressure during a weighing measurement.
This leads to the generation of additional chaotic forces depending
on the velocity pattern of the air flow in the close vicinity of the
pan of the balance. If the velocity pattern is not determined as well
as the impact of it on the balance, its influence on the measurand
is not known. Or, vibrations cause forces acting on the balance due
to accelerations. But if accelerations are not measured and additionally,
their impact are not known, the influence of them on the measurand
is not known either.

Therefore, the fundamental parameters can be distinguished between
known and unknown quantities such as,

\begin{equation}
x_{i}=g_{i}\left(\varepsilon_{1},\varepsilon_{2},..,\varepsilon_{l},h_{1},h_{2},..,h_{j}\right)
\end{equation}

where $h_{j}$ are ``hidden'' parameters. Thus the measurand is
given by,

\begin{equation}
y=f\left(g_{1}\left(\varepsilon_{1},\varepsilon_{2},..,\varepsilon_{l},h_{1},h_{2},..,h_{j}\right),...,g_{n}\left(\varepsilon_{1},\varepsilon_{2},..,\varepsilon_{l},h_{1},h_{2},..,h_{j}\right)\right)
\end{equation}

Consequently, the measurand becomes a function of fundamental known
and hidden parameters,

\begin{equation}
y=\widetilde{f}\left(\varepsilon_{1},\varepsilon_{2},..,\varepsilon_{l},h_{1},h_{2},..,h_{j}\right)
\end{equation}

\section{Origin of apparent random fluctuations of the Measurand}

Let us first assume that hidden fundamental parameters do not exist
and consequently the system is fully described by a well-known set
of equations. That means, that all fundamental parameters and their
impact on the behavior of the measurand are known. Hence the measurand
is given by,

\begin{equation}
y=\widetilde{f}\left(\varepsilon_{1},\varepsilon_{2},..,\varepsilon_{l}\right)\label{eq:2.1}
\end{equation}

Thus, at time $t=t_{0}$, the measurand is given by,

\begin{equation}
y\left(t_{0}\right)=\widetilde{f}\left(\left.\varepsilon_{1}\right|_{t_{0}},\left.\varepsilon_{2}\right|_{t_{0}},..,\left.\varepsilon_{l}\right|_{t_{0}}\right)
\end{equation}

at the time $t=t_{0}+\Delta t$, the measurand is given by,

\begin{equation}
y\left(t_{0}+\Delta t\right)=\widetilde{f}\left(\left.\varepsilon_{1}\right|_{t_{0}+\Delta t},\left.\varepsilon_{2}\right|_{t_{0}+\Delta t},..,\left.\varepsilon_{l}\right|_{t_{0}+\Delta t}\right)
\end{equation}

If the values of all known fundamental parameter at time $t=t_{0}+\Delta t$,
are equal to the values at time $t=t_{0}$, the value of the measurand
would be the same,

\begin{equation}
\left.\begin{array}{c}
\left.\varepsilon_{1}\right|_{t_{0}+\Delta t}=\left.\varepsilon_{1}\right|_{t_{0}}\\
\left.\varepsilon_{2}\right|_{t_{0}+\Delta t}=\left.\varepsilon_{2}\right|_{t_{0}}\\
\vdots\\
\left.\varepsilon_{l}\right|_{t_{0}+\Delta t}=\left.\varepsilon_{l}\right|_{t_{0}}
\end{array}\right\} \Rightarrow y\left(t_{0}+\Delta t\right)=y\left(t_{0}\right)\label{eq:2.4}
\end{equation}

Consequently the system is fully deterministic in that case.

Now let us assume that hidden fundamental parameters exist. The measurand
in that case is given by,

\begin{equation}
y=\widetilde{f}\left(\varepsilon_{1},\varepsilon_{2},..,\varepsilon_{l},h_{1},h_{2},..,h_{j}\right)
\end{equation}

Again, at time $t=t_{0}$, the measurand is given by,

\begin{equation}
y\left(t_{0}\right)=\widetilde{f}\left(\left.\varepsilon_{1}\right|_{t_{0}},\left.\varepsilon_{2}\right|_{t_{0}},..,\left.\varepsilon_{l}\right|_{t_{0}},\left.h_{1}\right|_{t_{0}},\left.h_{2}\right|_{t_{0}},..,\left.h_{j}\right|_{t_{0}}\right)
\end{equation}

and at time $t=t_{0}+\Delta t$, the measurand is given by,

\begin{equation}
y\left(t_{0}\right)=\widetilde{f}\left(\left.\varepsilon_{1}\right|_{t_{0}+\Delta t_{0}},\left.\varepsilon_{2}\right|_{t_{0}+\Delta t_{0}},..,\left.\varepsilon_{l}\right|_{t_{0}+\Delta t_{0}},\left.h_{1}\right|_{t_{0}+\Delta t_{0}},\left.h_{2}\right|_{t_{0}+\Delta t_{0}},..,\left.h_{j}\right|_{t_{0}+\Delta t_{0}}\right)
\end{equation}

However, since the impact of hidden parameters can not be evaluated
leads to the fact, that although in the case that the values of the
known parameters at both times are equal, the values of the measurand
at both times are not necessarily equal,

\begin{equation}
\left.\begin{array}{c}
\left.\varepsilon_{1}\right|_{t_{0}+\Delta t}=\left.\varepsilon_{1}\right|_{t_{0}}\\
\left.\varepsilon_{2}\right|_{t_{0}+\Delta t}=\left.\varepsilon_{2}\right|_{t_{0}}\\
\vdots\\
\left.\varepsilon_{l}\right|_{t_{0}+\Delta t}=\left.\varepsilon_{l}\right|_{t_{0}}
\end{array}\right\} \Rightarrow\begin{cases}
y\left(t_{0}+\Delta t\right)=y\left(t_{0}\right) & \left\{ \begin{array}{c}
\left.h_{1}\right|_{t_{0}+\Delta t}=\left.h_{1}\right|_{t_{0}}\\
\left.h_{2}\right|_{t_{0}+\Delta t}=\left.h_{2}\right|_{t_{0}}\\
\vdots\\
\left.h_{l}\right|_{t_{0}+\Delta t}=\left.h_{l}\right|_{t_{0}}
\end{array}\right.\\
y\left(t_{0}+\Delta t\right)\neq y\left(t_{0}\right) & \left.h_{i}\right|_{t_{0}+\Delta t}\neq\left.h_{i}\right|_{t_{0}}
\end{cases}
\end{equation}

\begin{figure}
\includegraphics[width=15cm]{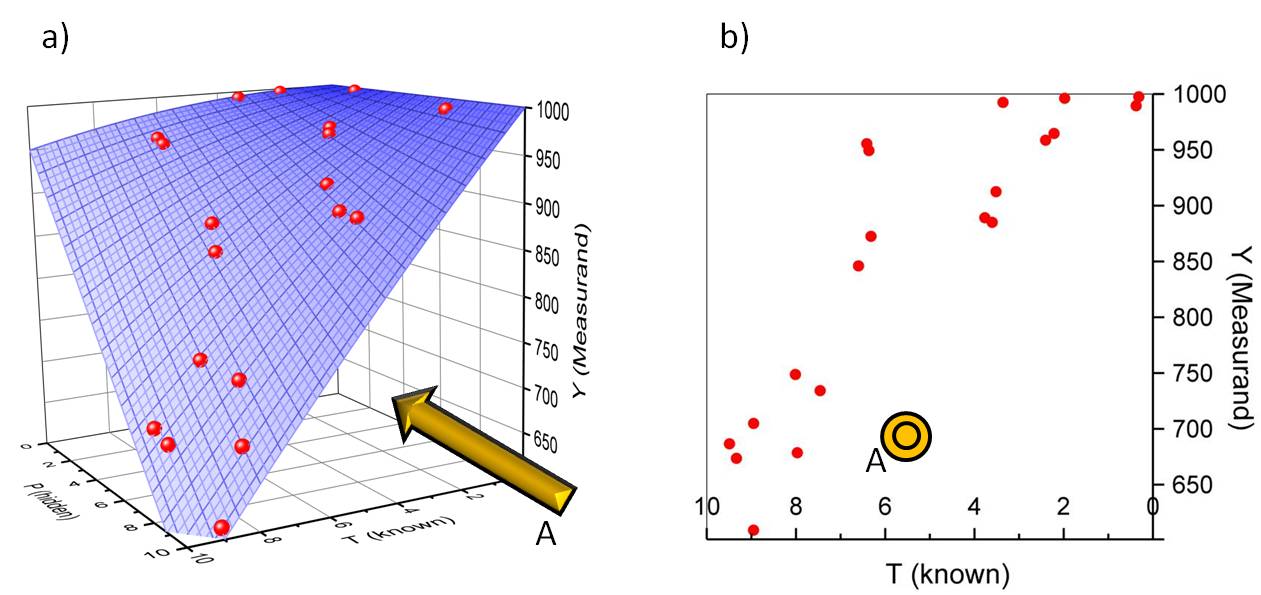}

\caption{\textit{\label{fig:3.1}Measurand as a function of hidden and known
parameter.} a) The measurand value is given for instance by $Y=-2T^{2}-4T\cdot P+1000$.
The parameter $T$ denotes the known temperature and $P$ is the unknown
pressure. The red balls indicate values of the measurand given by
specific values of the known and unknown variable. Here the impact
of $P$ on the measurand value is considered to be known. b) Due to
the fact, that the hidden parameter is not accessible, only the projection
(plane TY) of $Y$ is ``visible''. Thus, the measurand values apparently
scatter. }
\end{figure}

Strictly speaking, only in the case that the values of all input parameters
(known and hidden) are exactly the same, the value of the measurand
at both times would be equal. If only the value of one hidden parameter
is different at different times, the value of the measurand would
be different too. This leads to the fact, that for equal sets of known
input parameters the measurand can reach different values. Thus due
to lack of information of the system, it apparently behaves not necessarily
deterministic but rather reveals a stochastic behavior. This phenomenon
is depicted in Fig. (\ref{fig:3.1})

Nota bene, the stochastic behavior of the system is a consequence
of the existence of hidden parameters.

\section{Uncertainty of individual value of measurand}

Basically, the true value of a quantity is often not known. For instance,
considering hydrostatic weighing for the determination of liquid density.
Usually a solid body is immersed into the liquid and the apparent
loss of its mass (due to a lift force) is measured by using a balance.
The lift depends on the volume of the body which in turn depends on
temperature. But, the temperature of the body is not measured directly
since one avoids any generation of contact forces acting on the body.
Solely, the temperature of the fluid is measured in the vicinity of
the body. Thus, one can only estimate the true value of the body temperature.

To derive a relation for the uncertainty in that case, one calculates
the change of the value of the measurand for a small change of the
input parameters. This is given by,

\begin{equation}
\mathrm{d}y=\sum_{i}^{l}\left.\frac{\partial\widetilde{f}}{\partial\varepsilon_{i}}\right|_{\varepsilon_{1_{0}},\ldots,\varepsilon_{l_{0}}}\cdot\left(\varepsilon_{i}-\varepsilon_{i_{0}}\right)+\sum_{k}^{j}\left.\frac{\partial\widetilde{f}}{\partial h_{k}}\right|_{h_{1_{0}},\ldots h_{j_{0}}}\cdot\left(h_{k}-h_{k_{0}}\right)\label{eq:3.1}
\end{equation}

Now it is assumed that the true value of $\varepsilon_{i}$ and $h_{k}$
lies between $\left[\varepsilon_{i_{0}}-\Delta\varepsilon_{i},\varepsilon_{i_{0}}+\Delta\varepsilon_{i}\right]$
respectively $\left[h_{k_{0}}-\Delta h_{k},h_{k_{0}}+\Delta h_{k}\right]$
so that $\Delta\varepsilon_{i}$ and $\Delta h_{k}$ defines the range
in which we believe the true value lies with a specified likelihood.
Hence it is reasonable to chose $\varepsilon_{i}=\varepsilon_{i_{0}}+\Delta\varepsilon_{i}$,
and $h_{k}=h_{k_{0}}+\Delta h_{k}$ thus Eq. (\ref{eq:3.1}) becomes,

\begin{equation}
\mathrm{d}y=\sum_{i}^{l}\left.\frac{\partial\widetilde{f}}{\partial\varepsilon_{i}}\right|_{\varepsilon_{1_{0}},\ldots,\varepsilon_{l_{0}}}\Delta\varepsilon_{i}+\sum_{k}^{j}\left.\frac{\partial\widetilde{f}}{\partial h_{k}}\right|_{h_{1_{0}},\ldots h_{j_{0}}}\Delta h_{k}
\end{equation}

The value $\left(\mathrm{d}y\right)^{2}$ is a measure for the measurand
uncertainty ( $\left(\mathrm{d}y\right)^{2}$ instead of $\mathrm{d}y$
since the uncertainty should be positive). Thus,

\begin{equation}
\begin{aligned}u^{2}\left(y\right)=\left(\mathrm{d}y\right)^{2} & = &  & \sum_{i}^{l}\left(\left.\frac{\partial\widetilde{f}}{\partial\varepsilon_{i}}\right|_{\varepsilon_{1_{0}},\ldots,\varepsilon_{l_{0}}}\Delta\varepsilon_{i}\right)^{2}+2\sum_{i}^{l-1}\sum_{k=i+1}^{l}\left.\frac{\partial\widetilde{f}}{\partial\varepsilon_{i}}\right|_{\varepsilon_{1_{0}},\ldots,\varepsilon_{l_{0}}}\left.\frac{\partial\widetilde{f}}{\partial\varepsilon_{k}}\right|_{\varepsilon_{1_{0}},\ldots,\varepsilon_{l_{0}}}\Delta\varepsilon_{i}\Delta\varepsilon_{k}\\
 &  & + & \sum_{i}^{j}\left(\left.\frac{\partial\widetilde{f}}{\partial h_{k}}\right|_{h_{1_{0}},\ldots h_{j_{0}}}\Delta h_{i}\right)^{2}+2\sum_{i}^{j-1}\sum_{k=i+1}^{j}\left.\frac{\partial\widetilde{f}}{\partial h_{i}}\right|_{h_{1_{0}},\ldots h_{j_{0}}}\left.\frac{\partial\widetilde{f}}{\partial h_{k}}\right|_{h_{1_{0}},\ldots h_{j_{0}}}\Delta h_{i}\Delta h_{k}\\
 &  & + & 2\sum_{i}^{l}\sum_{k}^{j}\left.\frac{\partial\widetilde{f}}{\partial\varepsilon_{i}}\right|_{\varepsilon_{1_{0}},\ldots,\varepsilon_{l_{0}}}\left.\frac{\partial\widetilde{f}}{\partial h_{k}}\right|_{h_{1_{0}},\ldots h_{j_{0}}}\Delta\varepsilon_{i}\Delta h_{k}
\end{aligned}
\label{eq:3.2}
\end{equation}

Obviously, since the impact of hidden parameters can not be quantified,
the uncertainty of the value of the measurand can not be determined
either. However, a reasonable procedure to determine the uncertainty
is to consider the variance of the measurand at constant known fundamental
parameters. This is clear if we look on equation (\ref{eq:2.4}).
For constant known fundamental parameters the value of the measurand
is also constant. Hence, if any fluctuation (scatter) of the measurand
is observed at constant known parameters one can readily conclude
that this fluctuations must be caused by hidden parameters (Fig. \ref{fig:3.1}).

\begin{figure}
\includegraphics[width=15cm]{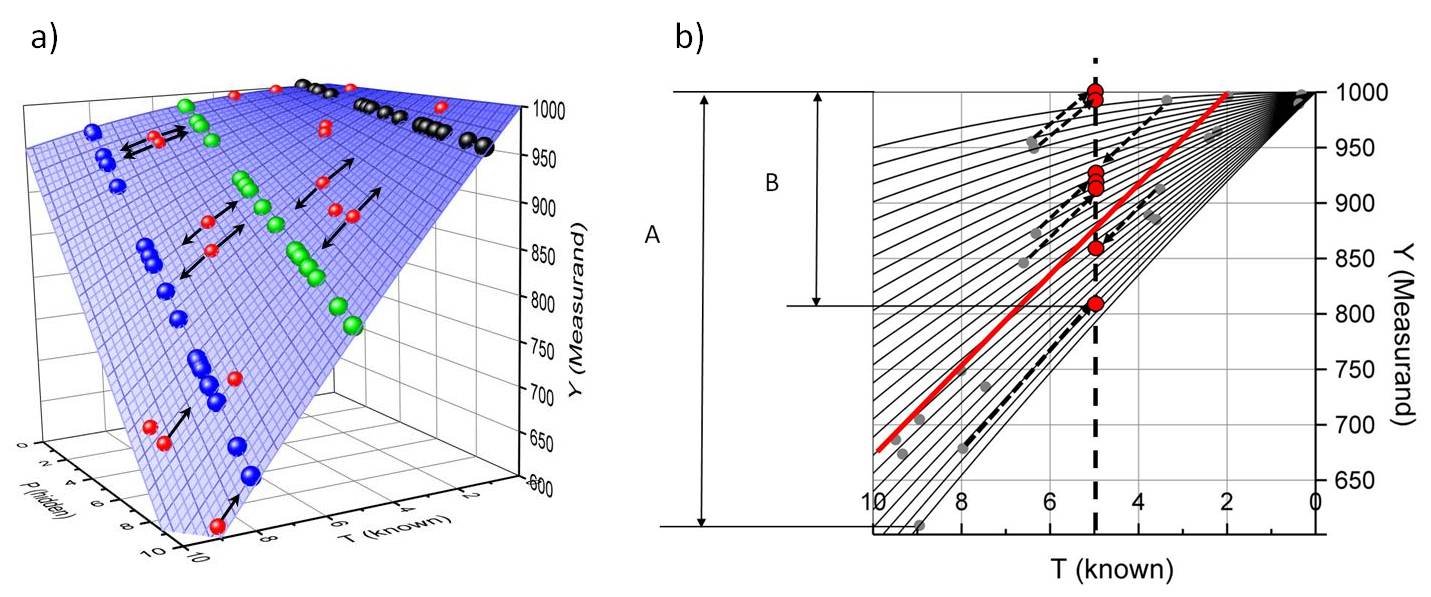}

\caption{\label{fig:3.3-1}\textit{Transformation of the measurand values}.
(a) In order to calculate the variance, the measurand values has to
be transformed according to the fundamental relationship $Y=-2T^{2}-4T\cdot P_{C}+1000$
where $P_{C}$ is equal to a given pressure value. (b) However, the
parameter $P$ is hidden. Thus in turn it is necessary to approximate
the temperature characteristics of the measurand $Y$. Thus the measurand
values have to be transformed according to a fit function to a specific
value of the known parameter $T$ (usually the mean value) (dashed
black line). Otherwise the variance would be overestimated due to
an over sized scatter interval (A). The red bold line shows the fit
function (in that case a linear fit was chosen). The red balls indicates
the transformed measurand values. It is also evident, that a linear
transformation (linear fit) is just a approximation. In fact, the
black solid lines depicts the functional relationship between the
measurand and the temperature at given preassures. It is clear, that
this functional relationship would be the best fit function to transform
every specific data point. But unfurtunatelly, $P$ is hidden, and
one only observes the situation depicted on the right figure without
any information of the true functional relationship between $T$ and
$Y$}
\end{figure}

Thus all values of the measurand must be transformed to the same set
of known parameters (Fig. \ref{fig:3.3-1}) . This can be achieved
by calculating a fit function of the measurand values. Thus the transformed
measurand values, $\gamma_{i}$, are given by,

\begin{equation}
\gamma_{i}=f:\,y\left(\varepsilon_{1_{i}},\varepsilon_{2_{i}},...,\varepsilon_{n_{i}}\right)\mapsto y(\varepsilon_{1_{i}}=A_{1},\varepsilon_{2_{i}}=A_{2},...,\varepsilon_{n_{i}}=A_{n})\label{eq:3.4-1}
\end{equation}

where the parameters $A_{n}$ are constants.

Nota bene, this fluctuation of the measurand value does not really
exist. They are quasi existing due to the lack of full information
of the system.

\begin{figure}
\includegraphics[width=15cm]{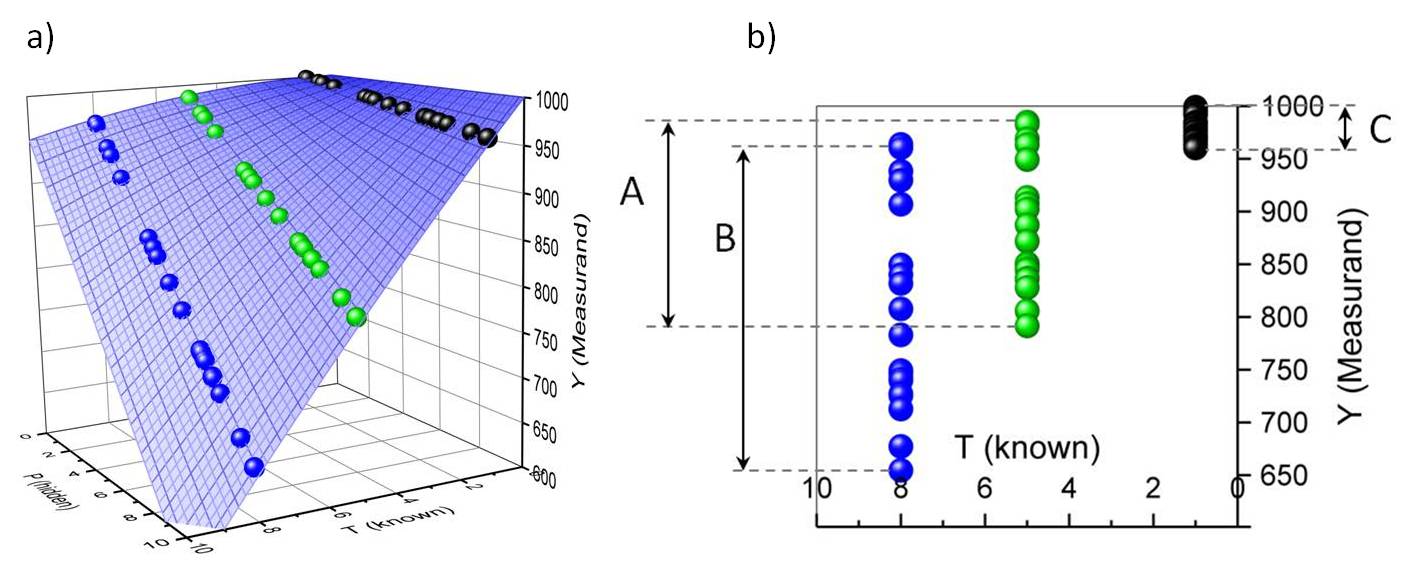}

\caption{\label{fig:3.2}\textit{Scatter of the measurand.} (a) At constant
known parameter $T$ the measurand exhibits a characteristic behavior
dependent on the hidden parameter $P$. (b) The measurand value apparently
shows an stochastic behavior in the accessible projection plane TY.
Thus, the measurand values scatters due to lack of information of
the system. As it is depicted, it is obvious that the scatter interval
(A,B,C) may depent on the known parameter $T$.}
\end{figure}

Hence, the individual uncertainty of the measurand value becomes,

\begin{equation}
\begin{aligned}u^{2}\left(y\right)=\left(\mathrm{d}y\right)^{2} & = &  & \overset{\mathrm{TYPE\,\,}\mathrm{B}}{\overbrace{\sum_{i}^{l}\left(\left.\frac{\partial\widetilde{f}}{\partial\varepsilon_{i}}\right|_{\varepsilon_{1_{0}},\ldots,\varepsilon_{l_{0}}}\Delta\varepsilon_{i}\right)^{2}+2\sum_{i}^{l-1}\sum_{k=i+1}^{l}\left.\frac{\partial\widetilde{f}}{\partial\varepsilon_{i}}\right|_{\varepsilon_{1_{0}},\ldots,\varepsilon_{l_{0}}}\left.\frac{\partial\widetilde{f}}{\partial\varepsilon_{k}}\right|_{\varepsilon_{1_{0}},\ldots,\varepsilon_{l_{0}}}\Delta\varepsilon_{i}\Delta\varepsilon_{k}}}\\
 &  & + & \underset{\mathrm{TYPE\,\,}\mathrm{A}}{\underbrace{Var\left(\gamma\right)}}
\end{aligned}
\label{eq:3.4}
\end{equation}

where the first term on the right hand side of Eq. (\ref{eq:3.4})
determines the Type B contribution and the second term determines
the Type A contribution to the overall uncertainty. Type B uncertainties
are calculated by deduction from an given joint probability density
function, $p\left(\varepsilon_{1},...,\varepsilon_{i},...\varepsilon_{N}\right)$
by \cite{key-3},

\begin{equation}
\left(\Delta\varepsilon_{i}\right)^{2}=\intop_{-\infty}^{+\infty}\mathrm{\left(\varepsilon_{i}-\left\langle \varepsilon_{i}\right\rangle \right)^{2}p\left(\varepsilon_{i}\right)d\varepsilon_{i}}=\left\langle \varepsilon_{i}^{2}\right\rangle -\left\langle \varepsilon_{i}\right\rangle ^{2}
\end{equation}

with,

\begin{equation}
p\left(\varepsilon_{i}\right)=\intop_{-\infty}^{+\infty}...\intop_{-\infty}^{+\infty}p\left(\varepsilon_{1},...,\varepsilon_{i},...\varepsilon_{N}\right)\mathrm{d}\varepsilon_{1}...\mathrm{d}\varepsilon_{i-1}\mathrm{d}\varepsilon_{i+1}...\mathrm{d}\varepsilon_{N}
\end{equation}

The Type B correlations in Eq. (\ref{eq:3.4}) are given by,

\begin{equation}
\Delta\varepsilon_{i}\Delta\varepsilon_{k}=\Delta\varepsilon_{i}^{2}=\intop_{-\infty}^{+\infty}\intop_{-\infty}^{+\infty}\mathrm{\left(\varepsilon_{i}-\left\langle \varepsilon_{i}\right\rangle \right)\left(\varepsilon_{k}-\left\langle \varepsilon_{k}\right\rangle \right)p\left(\varepsilon_{i},\varepsilon_{k}\right)d\varepsilon_{i}d\varepsilon_{k}}\label{eq:3.8-1}
\end{equation}

with,

\begin{equation}
p\left(\varepsilon_{i},\varepsilon_{k}\right)=\intop_{-\infty}^{+\infty}...\intop_{-\infty}^{+\infty}p\left(\varepsilon_{1},...,\varepsilon_{i},\varepsilon_{k},...\varepsilon_{N}\right)\mathrm{d}\varepsilon_{1}...\mathrm{d}\varepsilon_{i-1}\mathrm{d}\varepsilon_{i+1}...\mathrm{d}\varepsilon_{k-1}\mathrm{d}\varepsilon_{k+1}...\mathrm{d}\varepsilon_{N}
\end{equation}

Introducing a correlation coefficient, $C_{ik}$, for Typ B correlation,
Eq. (\ref{eq:3.8-1}) becomes,

\begin{equation}
\Delta\varepsilon_{i}\Delta\varepsilon_{k}=C_{ik}\sqrt{\left(\Delta\varepsilon_{i}\right)^{2}\left(\Delta\varepsilon_{k}\right)^{2}}
\end{equation}

whereas Type A contributions are calculated by induction via the variance
which is given by,

\begin{equation}
Var\left(\gamma\right)=\frac{\sum_{i}^{N}\left(\gamma_{i}-\overline{\gamma}\right)^{2}}{N-1}
\end{equation}

The mean value of the measurand is given by,

\begin{equation}
\overline{\gamma}=\frac{\sum_{i}^{N}\gamma_{i}}{N}\label{eq:3.8}
\end{equation}

It is important to emphasis that,

\begin{equation}
\begin{aligned}Var\left(\gamma\right) & \neq &  & \sum_{i}^{j}\left(\left.\frac{\partial\widetilde{f}}{\partial h_{i}}\right|_{h_{1_{0}},\ldots h_{j_{0}}}\Delta h_{i}\right)^{2}+2\sum_{i}^{j-1}\sum_{k=i+1}^{j}\left.\frac{\partial\widetilde{f}}{\partial h_{i}}\right|_{h_{1_{0}},\ldots h_{j_{0}}}\left.\frac{\partial\widetilde{f}}{\partial h_{k}}\right|_{h_{1_{0}},\ldots h_{j_{0}}}\Delta h_{i}\Delta h_{k}\\
 &  & + & 2\sum_{i}^{l}\sum_{k}^{j}\left.\frac{\partial\widetilde{f}}{\partial\varepsilon_{i}}\right|_{\varepsilon_{1_{0}},\ldots,\varepsilon_{l_{0}}}\left.\frac{\partial\widetilde{f}}{\partial h_{k}}\right|_{h_{1_{0}},\ldots h_{j_{0}}}\Delta\varepsilon_{i}\Delta h_{k}
\end{aligned}
\end{equation}

Depending on the magnitude of $\frac{\partial\widetilde{f}}{\partial h_{k}}\Delta h_{k}$
and on the stability (variation) of the hidden parameters the variance
could be,

\begin{equation}
\begin{aligned}Var\left(\gamma\right) & \geq &  & \sum_{i}^{j}\left(\left.\frac{\partial\widetilde{f}}{\partial h_{i}}\right|_{h_{1_{0}},\ldots h_{j_{0}}}\Delta h_{i}\right)^{2}+2\sum_{i}^{j-1}\sum_{k=i+1}^{j}\left.\frac{\partial\widetilde{f}}{\partial h_{i}}\right|_{h_{1_{0}},\ldots h_{j_{0}}}\left.\frac{\partial\widetilde{f}}{\partial h_{k}}\right|_{h_{1_{0}},\ldots h_{j_{0}}}\Delta h_{i}\Delta h_{k}\\
 &  & + & 2\sum_{i}^{l}\sum_{k}^{j}\left.\frac{\partial\widetilde{f}}{\partial\varepsilon_{i}}\right|_{\varepsilon_{1_{0}},\ldots,\varepsilon_{l_{0}}}\left.\frac{\partial\widetilde{f}}{\partial h_{k}}\right|_{h_{1_{0}},\ldots h_{j_{0}}}\Delta\varepsilon_{i}\Delta h_{k}
\end{aligned}
\end{equation}

or,

\begin{equation}
\begin{aligned}Var\left(\gamma\right) & \leq &  & \sum_{i}^{j}\left(\left.\frac{\partial\widetilde{f}}{\partial h_{i}}\right|_{h_{1_{0}},\ldots h_{j_{0}}}\Delta h_{i}\right)^{2}+2\sum_{i}^{j-1}\sum_{k=i+1}^{j}\left.\frac{\partial\widetilde{f}}{\partial h_{i}}\right|_{h_{1_{0}},\ldots h_{j_{0}}}\left.\frac{\partial\widetilde{f}}{\partial h_{k}}\right|_{h_{1_{0}},\ldots h_{j_{0}}}\Delta h_{i}\Delta h_{k}\\
 &  & + & 2\sum_{i}^{l}\sum_{k}^{j}\left.\frac{\partial\widetilde{f}}{\partial\varepsilon_{i}}\right|_{\varepsilon_{1_{0}},\ldots,\varepsilon_{l_{0}}}\left.\frac{\partial\widetilde{f}}{\partial h_{k}}\right|_{h_{1_{0}},\ldots h_{j_{0}}}\Delta\varepsilon_{i}\Delta h_{k}
\end{aligned}
\end{equation}

\begin{figure}
\includegraphics[width=15cm]{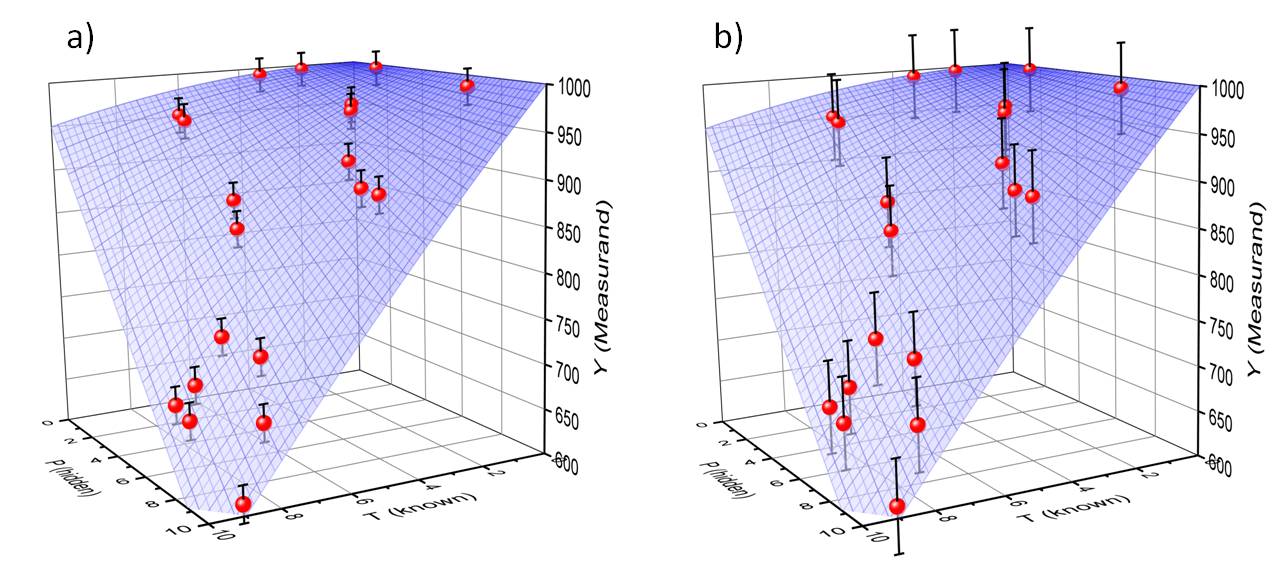}

\caption{\label{fig:3.2-1}\textit{Uncertainty of a single observation.} (a)
For a deterministic system (no hidden parameters) the uncertainty
is given by Eq. (\ref{eq:3.2}). (b) Due to lack of information the
uncertainty for a single measurand value may be overrated depending
on the scatter interval (see Fig. (\ref{fig:3.3-1}b).}
\end{figure}

Generally, in most cases (see Fig. \ref{fig:3.2}) due to lack of
information (existence of hidden parameters) the total uncertainty
for a single observation of the measurand value will be overrated
by applying statistical methods (Fig. \ref{fig:3.2-1}). It is just
a ``tool'' to account for uncertainties related to hidden parameters
.

Nota bene, for a non-linear relationship between the fundamental parameters
and the measurand, $y$, the Type B uncertainty according to Eq. (\ref{eq:3.4})
would give a wrong contribution to the overall uncertainty of the
measurand. In such a case a suitable procedure is given by the Monte
Carlo (MC) method to calculate the Type B uncertainty contribution.
According to this method, one calculates the measurand several times
where the fundamental parameters are picked from a probability density
distribution, $P_{\varepsilon_{n}}\left(\varepsilon_{n_{i}},\Delta\varepsilon_{n_{i}}\right)$
with expectation value, $\varepsilon_{n_{i}}$, and variance $\Delta\varepsilon_{n_{i}}$,
It is given by,

\begin{equation}
\widehat{y}_{ij}=\left(\widehat{\varepsilon}_{1_{ij}},\widehat{\varepsilon}_{2_{ij}},...,\widehat{\varepsilon}_{n_{ij}}\right)
\end{equation}

with,

\begin{equation}
\widehat{\varepsilon}_{n_{ij}}=P_{\varepsilon_{n}}\left(\varepsilon_{n_{i}},\Delta\varepsilon_{n_{i}}\right)_{j}
\end{equation}

The Type B uncertainty for an individual measurand value by applying
the MC method would then be given by,

\begin{equation}
u_{B}^{2}\left(y_{i}\right)=Var\left(\widehat{y}_{i}\right)=\frac{\sum_{j}^{M}\left(\widehat{y}_{ij}-\overline{\widehat{y}_{i}}\right)^{2}}{M-1}
\end{equation}

with,

\begin{equation}
\overline{\widehat{y}_{i}}=\frac{\sum_{j}^{M}\widehat{y}_{ij}}{M}
\end{equation}

where $M$ is the number of trials.

\section{The mean value of the measurand and its uncertainty}

According to the statements in sections (1), (2), and (3) it is inevitably
clear that the mean value of the measurand has to be evaluated for
a constant set of known fundamental parameters. Thus with Eq. (\ref{eq:3.4-1})it
is given by,

\begin{equation}
\overline{y}=\frac{\sum_{i}^{N}y_{i}(\varepsilon_{1}=A_{1},\varepsilon_{2}=A_{2},...,\varepsilon_{n}=A_{n})}{N}=\frac{\sum_{i}^{N}\gamma_{i}}{N}=\overline{\gamma}\label{eq:4.1}
\end{equation}

In principle, the constants $A_{n}$ are arbitrary but it is reasonable
to choose the mean values of the known parameters thus $A_{n}=\overline{\varepsilon_{n}}$.
Hence, the mean value of the measurand becomes,

\begin{equation}
\overline{y}=\frac{\sum_{i}^{N}y_{i}(\varepsilon_{1}=\overline{\varepsilon_{1}},\varepsilon_{2}=\overline{\varepsilon_{2}},...,\varepsilon_{n}=\overline{\varepsilon_{n}})}{N}\label{eq:4.2-1}
\end{equation}

Nota bene, if no hidden parameters would exist, each measurand $y_{i}$
for a constant set of known fundamental parameters would be given
according to Eq. (\ref{eq:2.1}) by,

\begin{equation}
y_{i}=y=\widetilde{f}\left(\varepsilon_{1}=A_{1},\varepsilon_{2}=A_{2},...,\varepsilon_{n}=A_{n}\right)
\end{equation}

With Eq. (\ref{eq:4.2-1}) this would lead to the fact, that,

\begin{equation}
\overline{y}=\widetilde{f}\left(\varepsilon_{1}=\overline{\varepsilon_{1}},\varepsilon_{2}=\overline{\varepsilon_{2}},...,\varepsilon_{n}=\overline{\varepsilon_{n}}\right)
\end{equation}

However, in that specific case, the measurand is totally deterministic
and the concept of a mean value and variance loses their meaning.
Furthermore, bear in mind, that the constants $A,B,C,...$ are totally
arbitrary.

The uncertainty of the mean is given with Eq. (\ref{eq:3.4}) choosing
$\varepsilon_{1_{0}}=\overline{\varepsilon_{1}},\ldots,\varepsilon_{l_{0}}=\overline{\varepsilon_{l}}$
by,

\begin{equation}
\begin{aligned}u^{2}\left(\overline{y}\right)=\left(\mathrm{d}\overline{y}\right)^{2} & = &  & \overset{\mathrm{TYPE\,\,}\mathrm{B}}{\overbrace{\sum_{i}^{l}\left(\left.\frac{\partial\widetilde{f}}{\partial\varepsilon_{i}}\right|_{\overline{\varepsilon_{1}},..,\overline{\varepsilon_{l}}}\Delta\varepsilon_{i}\right)^{2}+2\sum_{i}^{l-1}\sum_{k=i+1}^{l}\left.\frac{\partial\widetilde{f}}{\partial\varepsilon_{i}}\right|_{\overline{\varepsilon_{1}},..,\overline{\varepsilon_{l}}}\left.\frac{\partial\widetilde{f}}{\partial\varepsilon_{k}}\right|_{\overline{\varepsilon_{1}},..,\overline{\varepsilon_{l}}}\Delta\varepsilon_{i}\Delta\varepsilon_{k}}}\\
 &  & + & \underset{\mathrm{TYPE\,\,}\mathrm{A}}{\underbrace{\frac{1}{N^{2}}\left[\sum_{i=1}^{N}Var\left(\gamma\right)+2\sum_{i=1}^{N-1}\sum_{j=i+1}^{N}\mathit{COV}\left(\gamma_{i},\gamma_{j}\right)\right]}}
\end{aligned}
\end{equation}

where $\mathit{COV}\left(\gamma_{i},\gamma_{j}\right)$ accounts the
correlation between the transformed measurand value $\gamma_{i}$
and $\gamma_{j}$. 

The correlation term can be written as,

\begin{equation}
\mathit{COV}\left(\gamma_{i},\gamma_{j}\right)=r_{\gamma_{i},\gamma_{j}}\sqrt{Var\left(\gamma_{i}\right)}\sqrt{Var\left(\gamma_{j}\right)}
\end{equation}

where $r_{\gamma_{i},\gamma_{j}}$ is the correlation coefficient.
Since the values $\gamma_{i}$ and $\gamma_{j}$ belongs to the same
measurand, $r_{\gamma_{i},\gamma_{j}}$ is called auto correlation
coefficient. Thus, the Type A contribution to the overall uncertainty
of the mean of the measurand becomes,

\begin{equation}
u_{A}^{2}(\overline{y})=\frac{1}{N}\left[Var\left(\gamma\right)+\frac{2}{N}\sum_{i=1}^{N-1}\sum_{j=i+1}^{N}r_{y_{i},y_{j}}\sqrt{Var\left(\gamma_{i}\right)}\sqrt{Var\left(\gamma_{j}\right)}\right]\label{eq:4.3}
\end{equation}

\subsection{Case 1: $Var\left(\gamma_{i}\right)=Var\left(\gamma_{j}\right)=Var\left(\gamma\right)$,
$r_{\gamma_{i},\gamma_{j}}=r\protect\geq0$}

If we assume that all variances are equal as well as the auto correlation
coefficient between two transformed measurand value $\gamma_{i}$
and $\gamma_{j}$ , Eq. (\ref{eq:4.3}) becomes\footnote{$\sum_{i=1}^{N-1}\sum_{j=i+1}^{N}=\frac{1}{2}N\left(N-1\right)$},

\begin{equation}
u_{A}^{2}(\overline{y})=Var\left(\gamma\right)\left(r+\frac{1-r}{N}\right)
\end{equation}

It is evident, that for correlated system, the contribution of Type
A uncertainties to the overall uncertainty of the mean becomes in
the limit of $N\rightarrow\infty$,

\begin{equation}
\lim_{n\rightarrow\infty}u_{A}^{2}(\overline{y})=rVar\left(\gamma\right)
\end{equation}

In practice usually one encounters the fact, that the auto correlation
of the data is not considered in the total uncertainty of the mean.
Thus it is just often calculated by,

\begin{equation}
u_{A}^{2}(\overline{y})=\frac{Var\left(\gamma\right)}{N}\label{eq:4.6}
\end{equation}

Hence the uncertainty of the mean vanishes in the limit of $N\rightarrow\infty$.
Equation (\ref{eq:4.6}) should be utilized with care because it can
result in a strong underestimation of the uncertainty.

\subsection{Case 2: $Var\left(\gamma_{i}\right)\protect\neq Var\left(\gamma_{j}\right)$,
$r_{\gamma_{i},\gamma_{j}}\protect\neq r\protect\geq0$}

Usually the correlation coefficient between two random variables $x$
, $y$ is given by \cite{key-4},

\begin{equation}
r_{x,y}=\frac{\sum_{i}^{N}\left(x_{i}-\overline{x}\right)\left(y_{i}-\overline{y}\right)}{\sqrt{\sum_{i}^{N}\left(x_{i}-\overline{x}\right)^{2}\cdot\sum_{i}^{N}\left(y_{i}-\overline{y}\right)^{2}}}
\end{equation}

\begin{figure}
\includegraphics[width=10cm]{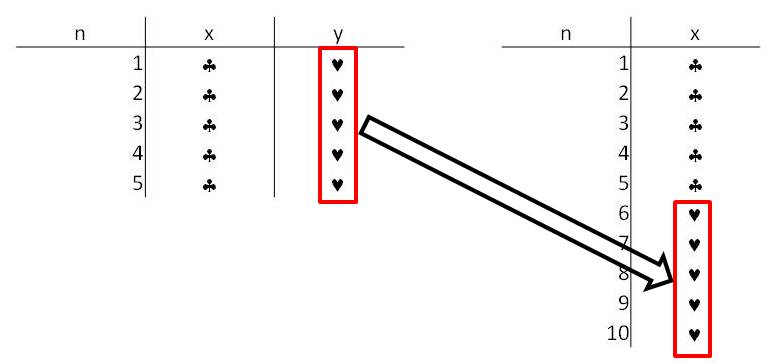}

\caption{\label{fig:3.2-1-1}\textit{ From correlation to auto correlation.}
Two separated random variables are merged to one data set.}
\end{figure}

Merging both variables (Fig. (\ref{fig:3.2-1-1})) the correlation
coefficient can be calculated as,

\begin{equation}
r_{x,y}=r_{x,x+5}=\frac{\sum_{i}^{\nicefrac{N}{2}}\left(x_{i}-\overline{x_{\leq5}}\right)\left(x_{i+5}-\overline{x_{>5}}\right)}{\sqrt{\sum_{i}^{\nicefrac{N}{2}}\left(x_{i}-\overline{x_{\leq5}}\right)^{2}\cdot\sum_{i}^{\nicefrac{N}{2}}\left(y_{i}-\overline{x_{>5}}\right)^{2}}}
\end{equation}

with the mean values given by,

\begin{equation}
\begin{aligned}\overline{x_{\leq5}}=\frac{1}{\nicefrac{N}{2}}\sum_{i=1}^{\nicefrac{N}{2}}x_{i} &  &  & \overline{x_{>5}}=\frac{1}{\nicefrac{N}{2}}\sum_{i=6}^{\nicefrac{N}{2}}x_{i}\end{aligned}
\end{equation}

Hence in general, the auto correlation coefficient is given by,

\begin{equation}
r_{\gamma_{i},\gamma_{j}}=r_{\gamma_{k},\gamma_{k+m}}=\frac{\sum_{i=0}^{N-(k+m)}\left(\gamma_{k+i}-\overline{\gamma_{k}}\right)\left(\gamma_{k+m+i}-\overline{\gamma_{k+m}}\right)}{\sqrt{\sum_{i=0}^{N-(k+m)}\left(\gamma_{k+i}-\overline{\gamma_{k}}\right)^{2}\cdot\sum_{i=0}^{N-(k+m)}\left(\gamma_{k+m+i}-\overline{\gamma_{k+m}}\right)^{2}}}
\end{equation}

with,

\begin{equation}
\begin{aligned}Var\left(\gamma_{k}\right)=\sum_{i=0}^{N-(k+m)}\left(\gamma_{k+i}-\overline{\gamma_{k}}\right)^{2}\\
Var\left(\gamma_{k+m}\right)=\sum_{i=0}^{N-(k+m)}\left(\gamma_{k+m+i}-\overline{\gamma_{k+m}}\right)
\end{aligned}
\end{equation}

\begin{figure}
\includegraphics[width=10cm]{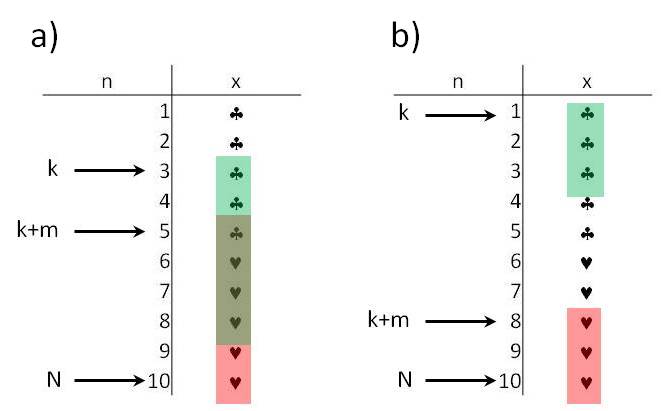}

\caption{\label{fig:3.2-1-1-1}\textit{Auto correlation.} (a) Auto correlation
coefficient calculation with a small step size $m$. (b) Large step
size leads to no overlap zone between the data.}
\end{figure}

For example, Fig. shows the auto correlation coefficient of a data
set with $N=150$.

\begin{figure}
\includegraphics[width=15cm]{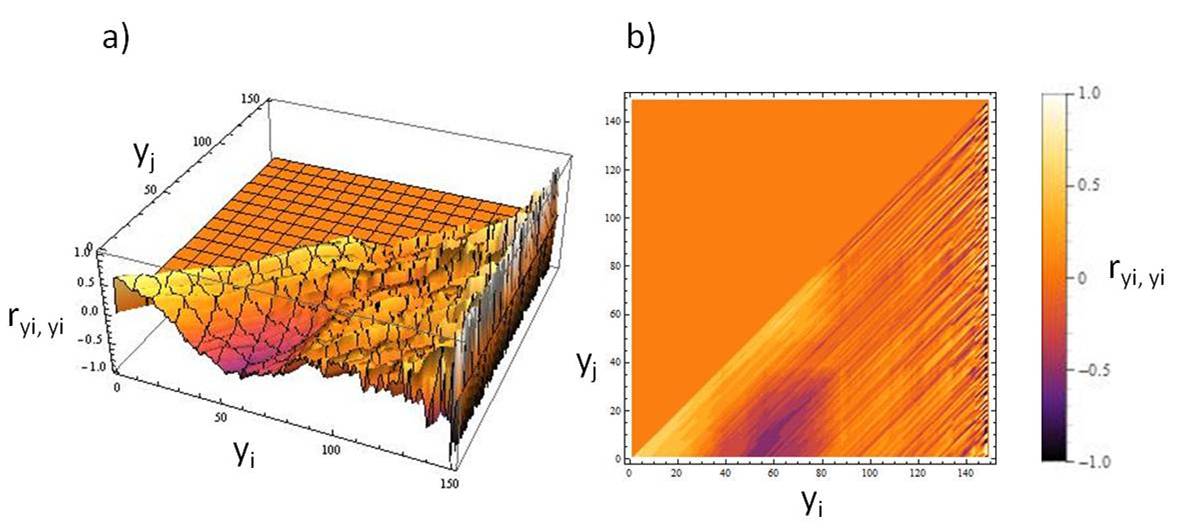}

\caption{\label{fig:3.2-1-1-1-1}\textit{Auto correlation coefficient for a
specific data set.} (a) Auto correlation coefficient. (b) Contour
plot of the auto correlation coefficient.}
\end{figure}

\end{document}